\newcommand{\jet}{\text{jet}}
\def\beq{\begin{equation}}
\def\eeq{\end{equation}}

\documentclass{cernrep}
\begin{document}
\title{Recent developments on parton-to-photon fragmentation functions}
%\title{Latest developments in parton-to-photon fragmentation functions}
%\author{A.N. Author and A.N. Other\thanks{On leave from another institue somewhere.}}
%\institute{Institute name in English, Town, Country}

\author{Tom Kaufmann${}^{\,1}$,  Asmita Mukherjee${}^{\,2}$, Werner Vogelsang${}^{\,1}$}
\institute{~\\[2mm]
${}^{1}$ Institute for Theoretical Physics, T\"ubingen University, 
Auf der Morgenstelle 14, \\ 72076 T\"ubingen, Germany\\[2mm]
${}^{2}$ Department of Physics, Indian Institute of Technology Bombay, \\
Powai, Mumbai 400076, India}

\begin{abstract}
We report on some recent developments concerning parton-to-photon fragmentation functions. We briefly summarize the 
main theoretical concepts on which the currently available extractions of these fragmentation functions rely: evolution
and the vector meson dominance model. We present comparisons of the available sets. We argue that the single-inclusive 
photon production process $pp\to\gamma X$ possesses only little sensitivity to the fragmentation 
contribution. Instead, the semi-inclusive process $pp\to(\jet\,\gamma)X$, for which the photon is observed inside a fully 
reconstructed jet and is part of the jet, is shown to offer much potential for providing valuable new information. 
\end{abstract}

%\keywords{CERN report; contribution; template; example.}

\maketitle

\section{State of the Art}

The production of photons with high transverse momentum $p_T$ in hadronic collisions is
of fundamental importance in today's particle and nuclear physics. Foremost, it may serve as a
tool for determining the gluon distributions of the scattering hadrons, thanks to the presence and 
dominance of the leading order (LO) Compton subprocess $qg\to \gamma q$ \cite{Vogelsang:1995bg,dEnterria:2012kvo}. 
Photons also provide sensitive probes of the medium produced in collisions of heavy ions, being able to 
traverse and escape the medium with little attenuation. Finally, photon signals play an important role
in studies of physics within and beyond the Standard Model, with the process 
$pp\to \gamma \gamma X$ through production and decay of a Higgs boson
and the indication of a 750 GeV diphoton excess seen by the 
ATLAS and CMS collaborations at the LHC in 2015 \cite{diphoton} arguably being among the most well-known examples.
Although the presumed ``bump" in the 2015 data disappeared with more 
statistics in a 2016 update \cite{no_diphoton}, the enormous interest
by the community (hundreds of papers tried to give a possible explanation of this excess \cite{750GeVrush}) 
demonstrated once again the importance of photon signals in the search for ``new physics"
beyond the Standard Model of particle physics.

As was discussed a long time ago~\cite{Owens:1986mp}, 
in perturbative high-$p_T$ processes there are two production mechanisms for photons. 
The photon may be directly produced in the hard scattering process through its ``pointlike'' QED coupling 
to a quark. Such contributions are usually referred to as ``direct". On the other hand, photons may also 
emerge in jet fragmentation, when a quark, antiquark, or gluon emerging from a QCD 
hard-scattering process fragments into a photon plus a number of hadrons. The need for introducing 
such a ``photon fragmentation'' contribution is physically motivated by the fact that the photon may result,
for example, from conversion of a high-$p_T$ $\rho$ meson.
Furthermore, at higher orders, the perturbative direct component contains divergencies from 
configurations where the photon and a (massless) final-state quark become collinear. These are long-distance 
contributions that naturally signify the need for non-perturbative fragmentation functions 
into which they can be absorbed.
Schematically, photon production cross sections may be written in a factorized form as
\beq\label{eq:xsec_gamma}
d\sigma^\gamma = d\hat{\sigma}^\gamma + \sum_c d\hat{\sigma}^c \otimes D_c^\gamma\,,
\eeq
where the sum runs over all partons (quarks, antiquarks and gluons). Note, that for processes with initial state hadrons,
the partonic cross sections in Eq.~\eqref{eq:xsec_gamma} have to be further convoluted with the respective PDFs.
%The first part in Eq.~\eqref{eq:xsec_gamma} is the called direct one, while 
%the second contribution is referred to as fragmentation" part. 
In general, the parton-to-photon fragmentation functions depend on the longitudinal momentum fraction $z$
which is transferred to the photon and on the factorization scale $\mu$: $D_c^\gamma \equiv D_c^\gamma(z,\mu^2)$. 
The non-perturbative functions $D_c^\gamma$ are assumed to be universal und thus may in principle
be extracted in the same manner as the parton-to-hadron FFs $D_c^h$ via fits to experimental data. 
While there has been much progress on parton-to-hadron FFs in the last years \cite{recent_hadron_FFs}, there have been
no new extractions of parton-to-photon FFs for about two decades, and our knowledge of these functions has
remained relatively poor. This is mostly due to the fact that the largest data set comes from
{\it single-inclusive} photon data in hadronic collisions, 
for which the dominant contribution comes from the direct (i.e. non-fragmentation) part. 
For the reaction $e^+e^-\to \gamma X$ fragmentation yields the dominant contribution; however,
here only a very sparse data set exists, and 
the amount of photon data and their precision does not match that of hadron production data.
The two most recent sets of photon FFs available are the ``Gl\"uck-Reya-Vogt" (GRV) set \cite{Gluck:1992zx} and the
``Bourhis-Fontannaz-Guillet" (BFG) set \cite{Bourhis:1997yu}. The latter is based on the somewhat older 
``Aurenche-Chiappetta-Fontannaz-Guillet-Pilon" (ACFGP) set \cite{Aurenche:1992yc}. 
We note,
that the BFG set actually consists of two sets of FFs, which mainly differ in the gluon-to-photon fragmentation function.

Besides universality, another key ingredient to extractions of FFs is evolution.
The presence of the direct part affects the evolution equations for photonic distributions.
Following \cite{Gluck:1992zx, Bourhis:1997yu}, the DGLAP-like evolution equations may be written as
\beq
\frac{d}{d\log\mu^2} D_i^\gamma(z,\mu^2) = \sum_j P_{ji}(z,\mu^2) \otimes D_j^\gamma(z,\mu^2)\,,
\eeq
where $i,j$ run over all parton flavors including the photon itself, i.e.~$i,j\in\{q_{i,j},\bar{q}_{i,j},g,\gamma\}$,
so that we also have a photon-to-photon FF $D_\gamma^\gamma$ and photon splitting functions. 
The symbol $\otimes$ denotes the standard convolution integral.
The evolution kernels, also called time-like splitting functions, are double series in the electromagnetic coupling $\alpha$
and the strong coupling $\alpha_s$,
\beq
P_{ij}(z,\mu^2) = \sum_{m,n} \left(\frac{\alpha(\mu^2)}{2\pi}\right)^m \left(\frac{\alpha_s(\mu^2)}{2\pi}\right)^n P_{ij}^{(m,n)}(z)\,.
\eeq
Usually, only the leading order in the electromagnetic coupling is considered, so that 
$D_\gamma^\gamma(z,\mu^2) = \delta(1-z)$. Furthermore, the running of $\alpha$ is neglected. The evolution 
equations then reduce to the frequently used inhomogeneous evolution equations
\beq\label{eq:inhom_evolution}
\frac{d}{d\log\mu^2} D_i^\gamma(z,\mu^2) = k_i^\gamma(z,\mu^2) + \sum_j P_{ji}(z,\mu^2) \otimes D_j^\gamma(z,\mu^2)\,,
\eeq
where now just $i,j\in\{q_{i,j},\bar{q}_{i,j},g\}$. The inhomogeneous term is given by
\beq
k_i^\gamma(z,\mu^2) = \frac{\alpha}{2\pi} \sum_n \left(\frac{\alpha_s(\mu^2)}{2\pi}\right)^n P^{(1,n)}_{\gamma i}(z)\,.
\eeq
Like for hadron FFs, these evolution equations are solved most conveniently in Mellin-$N$ space where the convolutions
turn into simple products. Furthermore, they usually are decomposed into the standard singlet and non-singlet
combinations, see, for instance, \cite{Gluck:1992zx, Bourhis:1997yu}. The full solution of the evolution equations 
\eqref{eq:inhom_evolution} is given by the sum of a particular solution to the inhomogeneous equation,
which can be computed in perturbation theory, and a general solution to the homogeneous equation, which contains 
the non-perturbative component. Schematically, we have
\beq\label{eq:evo_solution}
D_i^\gamma = D_i^{\gamma, \text{inhom}} + D_i^{\gamma, \text{hom}}\,.
\eeq
While the first part in Eq.~\eqref{eq:evo_solution} is perturbative, the second non-perturbative part has to be 
extracted from experiment or modeled. Ideally, one would prefer to extract this part from a clean reference process
without contamination of other non-perturbative functions (like PDFs), i.e.~from single-inclusive annihilation (SIA) 
$e^+e^-\to\gamma X$. However, only a very limited amount of data are available for this process, which furthermore 
have rather large uncertainties \cite{Ackerstaff:1997nha}. 
In view of this, the two most recent extractions of FFs have 
resorted to the vector meson dominance (VMD) model, for which one assumes that the fragmentation process is dominated by
conversion of vector mesons. Thus, the ansatz
\beq
D_i^{\gamma, \text{hom}}(z,\mu_0^2) = \alpha \sum_{v=\rho,\omega,\phi,\dots} C_v D_i^v(z,\mu_0^2)
\eeq
is used at the initial scale for the evolution, along with a vanishing inhomogeneous piece at the initial scale. 
Here, the sum runs over all vector mesons under consideration 
and the $D_i^v$ are the fragmentation functions into the corresponding vector mesons. As the FFs for 
the latter are rather poorly known as well, the GRV set adopts pion FFs for them instead. The 
BFG set uses $\rho$ data from ALEPH \cite{Buskulic:1995gm} and HRS \cite{Abachi:1989em} to constrain 
their VMD ansatz. We note that the recent paper \cite{Jezo:2016ypn} presented a detailed Monte-Carlo event generator 
study of the inclusive hadro-production of photons and vector mesons. By adding the $p_T$ spectra of the light vector 
mesons $(\rho,\omega,\phi)$, weighted with $\alpha$ divided by the individual vector-meson decay constants, one 
obtains a fairly good description of inclusive photon data in the low-$p_T$ region, which is the kinematic regime where the 
fragmentation process dominates over the direct part \cite{Klasen:2014xfa}. This observation supports the 
VMD ansatz for the nonperturbative part of photon FFs.

\begin{figure}[t]
\begin{center}
\includegraphics[width=\textwidth]{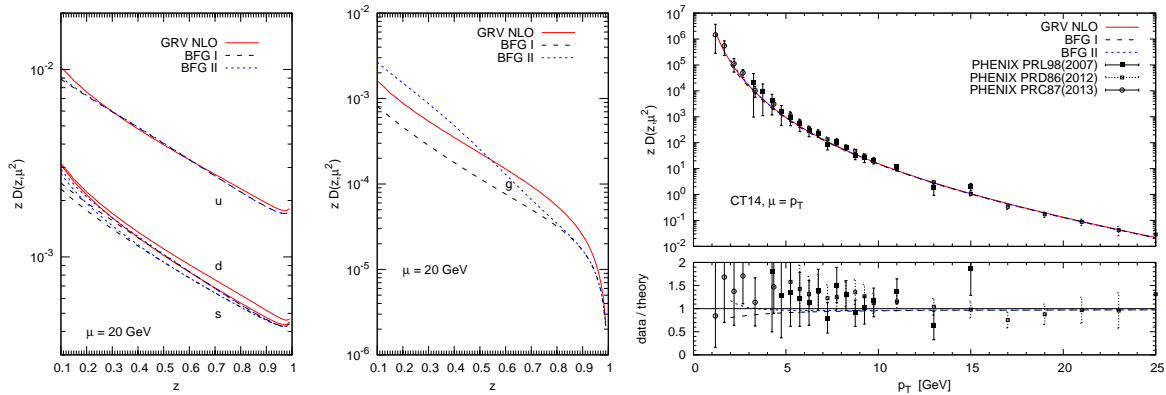}
\caption{Comparison of the different FF parametrizations \cite{Gluck:1992zx, Bourhis:1997yu}. In the left and middle panel we show
	              $z D_i^\gamma(z,\mu^2=(20\,\text{GeV})^2)$ for $i=u,d,s$ and $i=g$, respectively. In the right panel we 
	              compare to PHENIX data \cite{PHENIX_data} for $pp\to\gamma X$, where the lower part shows the ratio of the data 
	              and the theoretical NLO predictions for the two BFG sets with respect to the one based on GRV FFs.}
\label{fig:FF_compare}
\end{center}
\end{figure}

It is interesting to compare the different FF sets that are available~\cite{Vogelsang:1995bg,Klasen:2014xfa}. 
In the left and middle panels of Fig.~\ref{fig:FF_compare} we show $z D_i^\gamma(z,\mu^2 = (20\,\text{GeV})^2)$ for $i=u,d,s,g$,
as given by the GRV and the two BFG sets. While the quark FFs are rather similar, the gluon FF is quite different in each of 
the three sets. SIA data would not be expected to help discriminate among the various $D_i^g$ as the gluon FF enters only 
at NLO or via evolution. In order to see whether single-inclusive photon production in hadronic collisions, $pp\to \gamma X$, 
is more promising, we compare in the right panel of Fig.~\ref{fig:FF_compare} the theoretical cross section at NLO
(using the code of Ref.~\cite{Gordon:1993qc}) with data from PHENIX \cite{PHENIX_data}. 
As can be seen, the different FF sets yield very similar results. 
Compared to the experimental uncertainties, the difference in the FF sets
is negligible. Even though the gluon FF is very different and channels with {\it initial} gluons dominate, 
the inclusive photon production is apparently not really sensitive to the gluon FF. 
We note that this observation has been made in previous literature for various collider and 
fixed target settings; see, for instance, \cite{Vogelsang:1995bg,Klasen:2014xfa,theo_exp_others}.
We stress that the fact that the presently available sets of photon FFs yield similar cross section predictions 
does not imply that the fragmentation contribution to photon production is under satsifactory control.
For instance, there is arguably a much larger uncertainty in the $u$-quark FF than suggested by the 
curves shown in Fig.~\ref{fig:FF_compare}. 

These observations also have ramifications for photon signals in $pA$ and, especially, $AA$ collisions. For
the latter, photons are used in studies of the quark-gluon plasma (QGP). While photons produced directly
in the collision will traverse the medium with only little attenuation, the fragmentation photons will 
originate from partons that suffered energy loss in the medium. To assess this effect properly,
good understanding of the ``vacuum'' photon fragmentation functions is essential.

We finally note that in $pp$ collider experiments usually a photon isolation cut is introduced in order to 
suppress the large background from $\pi^0\to\gamma\gamma$ decay. The idea is to center
a cone around the final state photon and to demand that the hadronic energy fraction inside this cone be 
less than a certain amount $\epsilon$. Such isolation cuts also suppress the photon fragmentation contribution 
\cite{dEnterria:2012kvo,Gluck:1994iz}, since they confine the fragmentation contribution to large values of $z$.
This, however, introduces further uncertainties since the FFs are completely unconstrained in the region of large 
$z$ and since this region is much harder to treat theoretically.  For instance, large logarithmic $\log(1-z)$ contributions 
arise here in the evolution kernels and partonic cross sections~\cite{Catani:2002ny}. This especially affects the inhomogeneous part
which does not vanish for $z\to 1$~\cite{Gluck:1992zx}. Moreover, the isolation procedure also introduces logarithmic 
contributions in the energy fraction $\log\epsilon$ \cite{Catani:2002ny} and the cone opening 
\cite{Gordon:1994km,Catani:2013oma}. Any experimental observable that can provide direct information
on photon fragmentation at high $z$ will provide valuable insights into these questions. 
%%%%%%%%%%%%%%%%%%%%%%%%%%%%%%%%%%%%%%%%%%%%%%%%%%%%%%%%%%
%%%%%%%%%%%%%%%%%%%%%%%%%%%%%%%%%%%%%%%%%%%%%%%%%%%%%%%%%%
\section{Photon-in-jet production}
As we have seen, neither SIA nor inclusive photon production in $pp$ collisions are able to provide detailed access to the
fragmentation contribution. It is thus important to identify new observables that are able to yield new and complementary information
on photon FFs. One such observable was introduced in Ref.~\cite{Kaufmann:2016nux}, where the process $pp\to(\jet\,\gamma)X$
was proposed, for which a photon is observed in the final state inside a fully reconstructed jet, as part of the jet. 
Previous related studies \cite{jeth_theory} for the case of final-state hadrons had already established that such
``fragmentation-inside-jets'' observables may be reliably computed using factorization and perturbative-QCD techniques 
and may indeed give information on FFs. 
Recently, an extraction of $D^*$-meson FFs has been performed~\cite{ref:ourDstar}, including, for the first time, $D^*$-in-jet data.
It was shown that the in-jet data actually are able to give valuable constraints on, especially, the gluon FF. 
We now briefly present some of our results in~\cite{Kaufmann:2016nux}.

The cross section is calculated differential in the transverse momentum and rapidity of the jet, $p_T^\jet$ and $\eta^\jet$,
respectively, and the photon-jet momentum correlation variable
\beq
z_\gamma \equiv \frac{p_T^\gamma}{p_T^\jet}\,.
\eeq
The partonic cross sections are calculated analytically in the framework of the ``narrow jet approximation" (NJA). In the
NJA, the jet is assumed to be relatively narrow, in the sense that the jet parameter $R$ (we have in mind the widely used
anti-$k_T$ jet algorithm \cite{ Cacciari:2008gp} here) is rather small, $R \ll 1$. Thus, contributions of the order $\mathcal{O}(R^2)$
are neglected throughout the calculation. It was shown in Refs.~\cite{NJA} for inclusive jet production that the NJA works well
out to rather large values of $R\lesssim 0.7$.
We stress that the observable we have in mind is different from the ``away-side'' photon-jet correlations 
considered in Ref.~\cite{Belghobsi:2009hx} and provides a kinematically simpler and more direct access to the 
$D_c^\gamma$. 
%In particular, in our case it is natural to divide the cross section for $pp \to (\jet \gamma)X$
%             by the single-inclusive jet one for $pp \to \jet X$, in which case many theoretical uncertainties 
%             related to the choice of parton distributions or (initial-state) factorization scale will cancel out.
%
The main asset of the process $pp\to(\jet\,\gamma)X$ is that at LO the cross section is {\it directly proportional} to the FFs probed
at $z=z_\gamma$:
\beq\label{eq:jetgamma_LO}
\left.\frac{d\sigma^{pp\to(\jet\,\gamma)X}}{dp_T^\jet d\eta^\jet dz_\gamma}\right|_{\text{LO}} \propto 
\sum_{\begin{subarray}{c} a,b,c\,\in\,\\ \{q,\bar{q},g,\gamma\}\end{subarray}} f_a \otimes f_b \otimes d\hat{\sigma}_{ab}^{c,\text{LO}} 
\times \left[\delta(1-z_\gamma)\delta_{c\gamma} + D_c^\gamma(z_\gamma,\mu^2)(1-\delta_{c\gamma}) \right]\,.
\eeq
The first part in the squared brackets of Eq.~\eqref{eq:jetgamma_LO} is the direct part while the second one is the fragmentation
contribution. By demanding $z_\gamma < 1$ to ensure that we have a hadronic jet around the photon, the direct part does not
contribute at LO and only the fragmentation contribution remains. 
The various FFs are weighted by appropriate combinations of PDFs and
partonic cross sections, which may be regarded as ``effective charges''. The structure of the cross section hence becomes similar
to that for $e^+e^-\to \gamma X$, but with the essential difference that also gluon-to-photon fragmentation contributes
at the lowest order. 
We finally note, that we have presented a detailed study of the $\pi^0\to\gamma\gamma$ background for processes involving
final state photons in Ref.~\cite{Kaufmann:2016nux}. We have addressed the two main background sources,
i.e.~when the two decay photons become collinear or when one of the two photons falls below the energy detection
threshold.

%%%%%%%%%%%%%%%%%%%%%%%%%%%%%%%%%%%%%%%%%%%%%%%%%%%%%%%%%%
\begin{figure}[ht]
\begin{center}
\includegraphics[width=\textwidth]{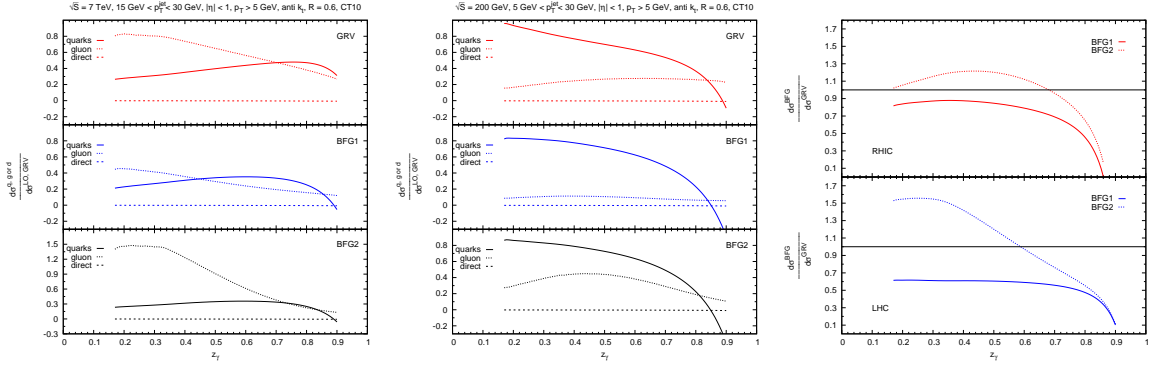}
\caption{The contribution of different subprocesses (gluon fragmentation, quark fragmentation and direct part) for 
	              kinematical setups corresponding to LHC (left panel) and RHIC (middle panel). In the right panel we show the
	              ratio of cross sections computed with the BFG sets with respect to that for GRV.}
\label{fig:jet+photon}
\end{center}
\end{figure}
%%%%%%%%%%%%%%%%%%%%%%%%%%%%%%%%%%%%%%%%%%%%%%%%%%%%%%%%%%
In Fig.~\ref{fig:jet+photon} we demonstrate the potential of the process $pp\to(\jet\,\gamma)X$ for providing
valuable and new information on the photon FFs. We show, for LHC and RHIC kinematics, 
the relative contributions of quark FFs, the gluon FF and the direct part for the three different FF parameterizations.
For better readability, we have normalized all results to the LO cross section computed with the LO GRV fragmentation set. 
We fist observe that the direct part is very small, being a pure NLO contribution. Furthermore, we see that
the gluon contribution is much larger for LHC compared to RHIC. This due to the large size of the contributions by
initial gluons at high center of mass energies.
In the right panel of Fig.~\ref{fig:jet+photon} we compare the full NLO cross sections for the different FF sets. We show the 
ratios of the cross sections computed with the BFG1 and BFG2 sets, relative to the cross section for the GRV FFs.
The potential of this process becomes visible, especially for the LHC setup where we find differences among the cross 
sections of up to $50\%$. This is in stark contrast to what we saw for $pp\to\gamma X$ (see the lower part of the right 
panel in Fig.~\ref{fig:FF_compare}), where over a large range in $p_T$ the cross sections for the different FF sets 
differed by less than $10\%$. We are hence optimistic that experimental data for $pp\to(\jet\,\gamma)X$ would allow 
one to distinguish between the different FF sets. 

\section{Conclusions}
We have presented the current state of the art for parton-to-photon fragmentation functions. While
precise knowledge of these functions is important for predictions for all observables with observed final-state photons, 
only little is actually known about them so far. The single-inclusive process $pp\to\gamma X$ has a 
dominant direct contribution, so that it is rather insensitive to the fragmentation one. For the SIA reaction
$e^+e^-\to\gamma X$, on the other hand, only a very sparse data set exists and it has no sensitivity to gluon 
fragmentation. In contrast to this, $pp\to(\jet\,\gamma)X$, for which a photon is observed in the final state inside
a fully reconstructed jet and is part of the jet, may provide direct and clean access to the parton-to-photon 
fragmentation functions, including the gluon one. We thus encourage experimental efforts to perform dedicated analyses
of this process. 

\section*{Acknowledgements}
The work of T.K.\ is supported by
the Bundesministerium f\"{u}r Bildung und Forschung (BMBF) under 
grant no.\ 05P15VTCA1. A.M.\ thanks the Alexander von Humboldt Foundation, Germany, for support through a Fellowship for Experienced Researchers.
%
%We wish to thank A.N. Colleague for enlightening comments on
%the present topic.

\end{document}